\RequirePackage{setspace}
\documentclass{ejb}
\usepackage{comment}
\usepackage{rotating}
\usepackage{xspace}
\usepackage{tikzscale}
\usepackage{filecontents}
\usepackage{tikz}
\usepackage{fancyvrb}
\usepackage{subfiles}
\usepackage{pgfplotstable}
\usepackage{longtable}
\usepackage{pgfplots}
\usepackage{hyperref}
\usepackage{helvet}
\usepackage[eulergreek]{sansmath}
\usepackage{graphicx}
\usepackage{subfig}
\usepackage{pdflscape}
\usepackage{multirow}
\usepackage{fancyhdr}
\usepackage{rotating}

\usepackage{amsthm}
\usepackage{graphicx}
\usepackage{multirow}
\usepackage{tabularx}
\usepackage{amssymb}
\usepackage{mathtools}
\usepackage[flushleft]{threeparttable}

\pgfplotsset{compat=1.3}

\fvset{formatcom=\color{black},fontsize=\scriptsize,frame=single,rulecolor
=\color{black},baselinestretch=1,framesep=5mm}

\DefineVerbatimEnvironment {GAMSCode*}{Verbatim}
{framesep=5mm,frame=none,fontsize=\scriptsize, baselinestretch=1}

\graphicspath{{./Figures/}}

\begin{document}

\title{Non-fundamental Home Bias in International Equity Markets}
\author{\small
Gyu Hyun Kim
\footnote{
Department of Economics, Iowa State University, Heady Hall, 
518 Farm House LN, Ames, IA 50011, USA. (email:
gyuhyun@iastate.edu).}
}

\maketitle


\begin{abstract}
This study investigates the relationship of the
equity home bias with 1) the country-level behavioral unfamiliarity,
and 2) the home-foreign return correlation. We set the hypotheses
that 1) unfamiliarity about foreign equities plays a role in the portfolio
set up and 2) the correlation of return on home and foreign equities
affects the equity home bias when there is a lack of information about
foreign equities. For the empirical analysis, the proportion of respondents
to the question ``How much do you trust? - People you meet for the first time"
is used as a proxy measure for country-specific unfamiliarity. Based on the
eleven developed countries for which such data are available, we implement
a feasible generalized linear squares (FGLS) method. Empirical results
suggest that country-specific unfamiliarity has a significant and
positive correlation with the equity home bias. When it comes to the correlation
of return between home and foreign equities, we identify that there
is a negative correlation with the equity home bias, which is against
our hypothesis. Moreover, an excess return on home equities compared to foreign ones 
is found to have a positive correlation with the equity home bias, 
which is consistent with the comparative statics 
only if foreign investors have a sufficiently higher risk aversion than domestic investors.
We check the robustness of our empirical analysis by fitting alternative specifications and use a log-transformed measure of the equity home bias, resulting in consistent results with ones with the original measure. 
\end{abstract}



\section{Introduction}\label{ehb-intro}

This paper investigates the role of behavioral bias in international
equity markets with information frictions. Investors
set an optimal portfolio that offers the highest return,
which is equivalent to the highest future wealth. When investors have
comprehensive information about foreign equities, they consider a
combination of sensible economic factors: returns, transaction costs,
risks, and market correlations. In this case, the equity home bias
is an outcome of transaction costs, such as differences in regulations
and operating costs of local offices for financial institutions. Nevertheless, in reality, 
investors indeed do not have comprehensive information about foreign financial assets.
In this case, a behavioral bias such as unfamiliarity
can work as part of the noise on the home country's information
about returns on foreign equities. For example, under a lack of information
about foreign equities, home investors may undervalue foreign equities,
though they reflect measurable transaction costs. In this study, we
investigate whether non-fundamental familiarity about foreign equities
is a significant determinant of the equity home bias.

\citet{french1991were} first pointed out the equity home bias by
observing that domestic equities comprise over 90\% of equity wealth
in the U.S and Japan. Since the 1990s, the equity home bias has been
a widely explored puzzle in international financial markets. \citet{lewis1999trying} and \citet{VANWINCOOP20101108} approach this puzzle with
hedge mechanisms. \citet{cooper2013equity} suggest that domestic investors
expect lower returns on foreign assets due to international transaction
costs. Current research analyzes
the equity home bias considering both institutional and behavioral
frameworks (\citet{lewis1999trying} and \citet{cooper2013equity}). 

\begin{figure}[ht]
        
        \caption{Share of Total Equity Portfolio in World Market Capitalization (Blue)
        and Share of Holdings of Domestic Equity in Total Equity Portfolio
        (Red)} \label{fig:ehb-share}
            \begin{threeparttable}
            
                \begin{centering}
                \includegraphics[scale=0.34]{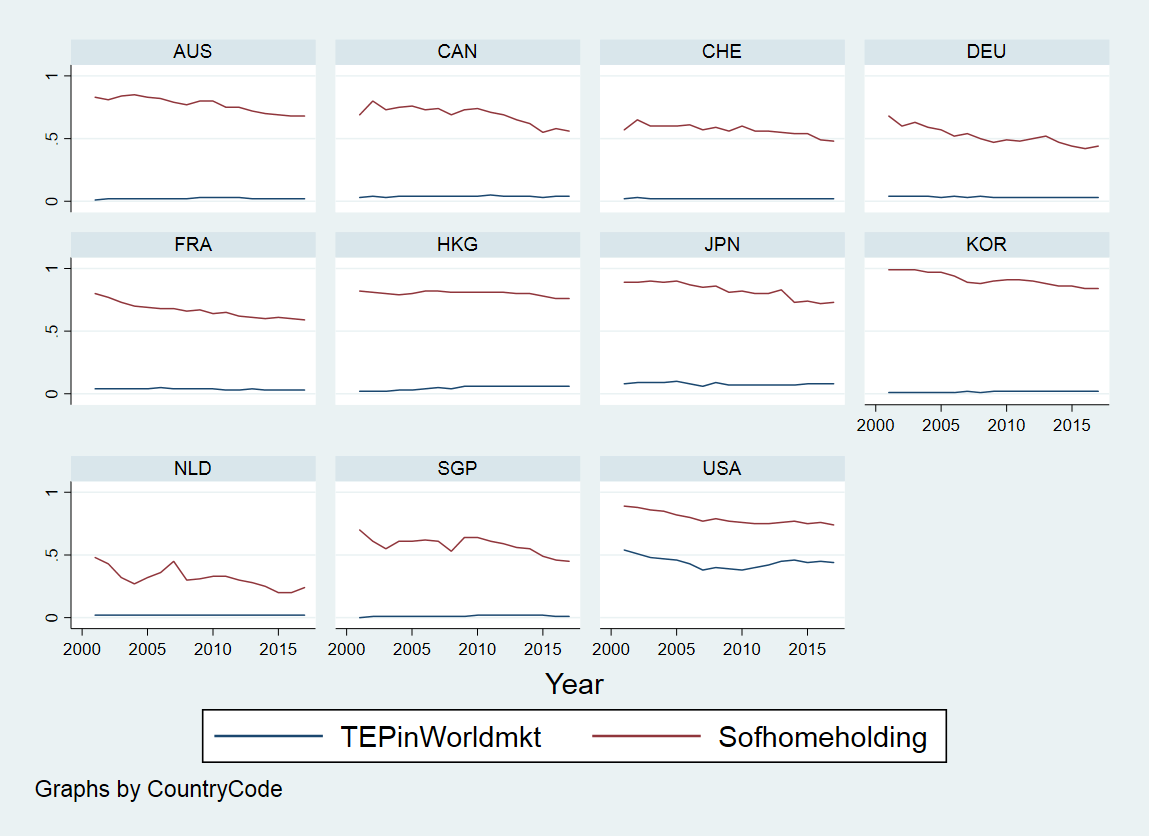}
                \par\end{centering}
                
            \begin{tablenotes}
            \item [Note:] \footnotesize{} Computed from IMF CPIS and World Bank data
            \end{tablenotes}
            \end{threeparttable}
            
            \vspace{0.5 in}
        \end{figure}

Furthermore, despite the high integration of the current international
finance market, equity home biases have been relatively stable over time. It is agreed
that integration in the international financial market leads to a
drastic reduction of the transaction costs incurred from different
languages and legal systems. For example, \citet{werner_1997}
and \citet{coeurdacier2013home} point out that this bias has been reduced
since the onset of financial market integration. However, as Figure
\ref{fig:ehb-share} shows, the share of domestic equity in the United States and Japan
is still over 70\%, and its trend is relatively stable, although the
global shares of total equity value in each country are less than
60\%: ranging from 38\% to 54\% in the U.S. and from 2\% to 3\%
in Japan between 2001 and 2014. 
    \footnote{From 2001 to 2014, the global shares of total equity
    value in each country are computed by dividing the total equity portfolio
    by the global total equity portfolio.} .
    
This study focuses on how two factors - namely, 1) information frictions about
foreign equities and 2) the correlation
between domestic and foreign return on equities - are related to the
equity home bias. Moreover, this work empirically shows that non-fundamental
unfamiliarity is a significant determinant of the equity home bias.
Based on a two-country model modified from that of \citet{chan2005determines}
and \citet{foad2011immigration}, the optimal equity shares of both domestic and foreign
equity are derived when we employ information frictions on foreign
equities. At the optimal share of the portfolio, this study examines
how the equity home bias depends on 1) the information frictions on
foreign equities and 2) the correlation between domestic and foreign
equity returns. For the empirical analysis, we assume that there are
information frictions about the foreign equities, and implement a feasible
generalized linear squares (FGLS) method. In the empirical analysis,
the country-level equity home bias is computed using data from the
World Bank and Coordinated Portfolio Invest Survey (CPIS). The country-level
proportion of respondents to the survey question, ``How
much you trust: People you meet for the first time?'',
in the World Value Survey (WVS) is exploited as a measure for the level
of non-fundamental unfamiliarity, which is an essential factor for
noise from the information frictions on foreign equities. In the last
part of this research, we conduct additional empirical analyses
using transformed measures of the equity home bias.

This work is innovative in two senses. Firstly, we use a novel measure
of personal behavior based on publicly available data. The question adopted from
the WVS data closely represents the country-level unfamiliarity. Using
this data, we support the relationship between country-specific unfamiliarity
and the equity home bias when home investors have limited information
about foreign equities. Previous studies, such as \citet{huberman2000home},
\citet{beugelsdijk2010cultural}, and \citet{chan2005determines}, investigate
the relationship between familiarity and the equity home bias using
proxies, such as languages and distance. Secondly, we investigate
behavioral factors and market correlation simultaneously. 

Comparative statics hypothesize an increase in the equity home bias under the information frictions. 
Moreover, a positive correlation between domestic and foreign equities leads to
a higher equity home bias, which explains that investors have less
incentives to buy foreign assets due to higher substitution. In the
empirical analysis, assuming that domestic financial markets have
information frictions about foreign equities, we verify that unfamiliarity
has a significant and positive correlation with the equity home bias.
Regarding the correlation of return between home and foreign equities,
we find a negative correlation with the equity home bias, which is
opposite to the result from the comparative statics. Additionally,
an excess return on home equities compared to foreign ones 
is found to have a positive association with the equity home bias, 
which is consistent with the comparative statics 
when foreign investors have a higher risk aversion than domestic investors. 
To check the robustness of our empirical analysis, 
we fit alternative estimation models, and use a log-transformed measure of the equity home bias,
which turns in consistent results with ones with the original measure.  

This paper proceeds as follows. In section \ref{ehb-literature}, related studies are
reviewed. Then, we set the theoretical model and provide analytic
solutions in section \ref{ehb-framework}. After the theoretical work, in section \ref{ehb-empirical},
the hypotheses are tested and the empirical results discussed. Finally,
in section \ref{ehb-conclusion}, the concluding remarks are provided.

\section{Related Literature}\label{ehb-literature}

Traditional studies approach the equity home bias with a transaction
cost, such as relevant fees incurred from cross-border financial
transactions (\citet{stultz_1981} and \citet{obstfeld2000six}). Beyond
transaction costs, there have been numerous studies exploring the
determinants of the equity home bias. The ability to hedge domestic
or exchange rate risks is considered as one of the most significant
determinants of the equity home bias 
(\citet{cooper1986costs}, \citet{cooper1994home},
\citet{mishra2011australia}, \citet{beck2008impact} and \citet{coeurdacier2013home}
). Moreover, asymmetric information 
(\citet{kang1997there}, \citet{ivkovic2005local}, \citet{foad2011immigration}) 
and corporate governance structures (\citet{gelos2005transparency}) 
are suggested as factors associated with the equity home bias.

The equity home bias is also explained as a result of the interaction
between financial markets. \citet{quinn2008century} and \citet{levy2014home}
argue that if markets' returns move in the same direction, then investors
have less incentives to use foreign equities for investment diversification.
Comovement of both domestic and foreign returns implies that there
is a high substitute relationship between domestic and foreign equities.
In this case, foreign equities are not attractive to hedge domestic
equity risk. \citet{levy2014home} show that home bias is proportional
to the average correlation between markets, and the proportion is
less than 1
    \footnote{Given costs, \citet{levy2014home} show that the home bias is
    proportional to $\dfrac{\rho}{1-\rho}$, where $\rho$ is the correlation 
    between market returns.}. 

Recently, there has been a consensus that the equity home bias is
an outcome of the combination of behavioral and institutional factors.
Using panel data from 38 countries between 2001 and 2010, 
\citet{bose2015education} show that improving financial literacy has a significant
impact on reducing the equity home bias, and its impact is amplified
in the less financially developed economies. 
\citet{beugelsdijk2010cultural} suggest that 
risk aversion and cultural differences
relate to unfamiliarity with foreign assets. \citet{huberman2000home} argues
that investors overestimate the risk of foreign assets because the
foreign assets are less familiar to domestic investors. Most of the
familiarity in those works is measured by official languages, the
distance between countries or cultural similarities 
(\citet{grinblatt2001distance}, \citet{chan2005determines} and \citet{berkel2007institutional}).
\citet{morse2011patriotism} introduce patriotism. 
Using the World Values Survey (WVS)
data, they show that patriotism can explain marginally five percent
of the equity home bias. \citet{barberis2003survey} introduce belief perseverance,
which is the investors' reluctance to look for evidence that refutes
the beliefs.

This study is in line with previous studies, which suggest that the
equity home bias arises from a combination of socio-behavioral features,
asymmetric information, and financial markets' interactions. This
work is in the context of the results from \citet{huberman2000home}, \citet{foad2010familiarity}, and \citet{foad2011immigration} in the sense that investors tend to avoid foreign equities
because unfamiliarity plays a role when there is a lack of information
about foreign equities. Since this work considers the comovement of
domestic-foreign market returns, it is also related to the study by
\citet{levy2014home}. 

Compared to the previous literature, this work has two distinguishable
points. Firstly, it utilizes a novel measure of personal behavior,
unfamiliarity, from publicly available data. Secondly, it analyzes 
both the behavioral factor and market correlation simultaneously.

\section{Theoretical Framework}\label{ehb-framework}

\citet{lewis1999trying} and \citet{foad2011immigration} suggest a mean-variance framework to analyze
the equity home bias under information frictions on foreign financial
markets. While sharing the spirit of the international capital asset
pricing model (ICAPM) of \citet{foad2011immigration}, this work sets the model without
inflation and exchange rate. For simplicity, we introduce a mean-variance
framework instead of a constant relative risk aversion model (CRRA).
Furthermore, to concentrate on the role of unfamiliarity in information
frictions about foreign equities, this work does not consider cross-border
labor movements such as immigration, which are incorporated in \citet{foad2011immigration}.

Following \citet{ljungqvist2018recursive}, we can derive a mean-variance
utility from an exponential utility function and a normally distributed
return on a portfolio 
    \footnote{We can also get a mean-variance utility from a second-order approximation
    of an expected utility function with respect to the level of future wealth.}. 
    
In Appendix \ref{ehb-appenA}, we show how maximizing the expected utility function
is equivalent to maximizing the mean-variance utility, which is a linear
function of mean and variance of return on a portfolio. Combining
the information set, $I_{h}$ for home country and $I_{f}$ for foreign
country, and the home investor's share of both domestic and foreign
equity, $w_{h}^{h}$ and $w_{h}^{f}=1-w_{h}^{h}$, we can define the
home investor's portfolio expected return and variance as:

    \begin{gather} \label{ehb-eq1}
    E\left(R_{h}^{P}|\,I_{h}\right)  =w_{h}^{h}\left[E\left(R^{h}\,|\,I_{h}\right)-c_{h}^{h}\right]+\left(1-w_{h}^{h}\right)\left[E\left(R^{f}\,|\,I_{h}\right)-c_{h}^{f}\right] \\
    (\sigma_{h}^{P})^{2}|\,I_{h} =\left(w_{h}^{h}\right)^{2}var\left(R^{h}\,|\,I_{h}\right)+\left(1-w_{h}^{h}\right)^{2}var\left(R^{f}\,|\,I_{h}\right)+ \nonumber \\ 2\left(w_{h}^{h}\right)\left(1-w_{h}^{h}\right)cov\left(R^{h},\,R^{f}\,|\,I_{h}\right) \nonumber. 
    \end{gather}

When it is assumed that short leverage is not allowed, $w_{h}^{h}$
is the non-negative domestic portfolio share of home country's total
investment, while $w_{h}^{f}$ is the foreign portfolio share of domestic
portfolio
    \footnote{A country might have unfamiliarity and information frictions that are
    differentiated across countries. For example, from the perspective of
    country $A$, the degree of unfamiliarity and information frictions
    about equity of country $B$ are not equal to those of country C.
    For simplicity, we assume that a foreign asset is a combination of
    multiple country-specific portfolios, and it is given. By doing so,
    we can specify unfamiliarity and information frictions, focusing on
    the local investor's point of view.}
    \footnote{Since we view unfamiliarity about foreign equities as a non-fundamental
    feature on which an investor depends when he/she has information frictions,
    we do not explicitly incorporate a degree of unfamiliarity into the
    framework in this study.}. 
Also, $R^{h}$ and $R^{f}$ are the expected return of each domestic
and foreign equity. $c_{h}^{h}\;\left(c_{h}^{f}\right)$ is defined
as the cost of home investors holding domestic (foreign) equity. The
objective function of the home investor in mean-variance form is:

    \begin{gather} \label{ehb-eq2}
    Max_{w_{h}^{h}}\;w_{h}^{h}\left[E\left(R^{h}\,|\,I_{h}\right)-c_{h}^{h}\right]+\left(1-w_{h}^{h}\right)\left[E\left(R^{f}\,|\,I_{h}\right)-c_{h}^{f}\right]\\
    -\dfrac{\lambda_{h}}{2}\left[(\sigma_{h}^{P})^{2}|\,I_{h}\right]\nonumber
    \end{gather}
where $\lambda_{h}\;(\lambda_{f})$ is home (foreign) investor's
degree of risk aversion. In order to focus on the issue of our interest, we do
not incorporate currency differences which are associated with
exchange rate risk. Since the mean-variance utility is equivalent
to the expected utility of an investor's future wealth from the portfolio,
we can define consumption of a representative investor as the expected
future wealth of the portfolio in this framework. Given a positive
optimal portfolio share, it implies a positive relationship between
domestic equity return and consumption.

From the home investors' perspective, there are information frictions
on foreign equities. Likewise, foreign investors do not have complete
information on home equities. Especially, according to \citet{foad2011immigration},
given the information set of $I_{h}$, home investors consider a degree
of uncertainty on foreign equities quantified by $\eta_{f}$. Also,
foreign investors with $I_{f}$ face risk about home equities in the
amount $\eta_{h}$. Denoting $\eta_{h}\;\left(\eta_{f}\right)$ as
a variance of return on home (foreign) equity from the perspective
of foreign (home) investors with information set $I_{f}\;(I_{h})$,
we can describe the following relationships when we assume $u_{f}$
$(u_{h})$ does not have a correlation with $R^{h}\,|\,I_{f}$ $(R^{f}\,|\,I_{h})$:

    \begin{gather*}
     R^{h}\,|\,I_{f}=R^{h}\,|\,I_{h}+u_{f}\;where\;u_{f}\sim\left(0,\,\eta_{f}\right)\\
     E\left[R^{h}\,|\,I_{f}\right]=E\left[R^{h}\,|\,I_{h}\right]\equiv\mu^{h}\\
     var\left(R^{h}\,|\,I_{f}\right)=var\left(R^{h}\,|\,I_{h}\right)+\eta_{f}\equiv\left(\sigma^{h}\right)^{2}+\eta_{f}\\\\
     R^{f}\,|\,I_{h}=R^{f}\,|\,I_{f}+u_{h}\;where\;u_{h}\sim\left(0,\,\eta_{h}\right)\\
     E\left[R^{f}\,|\,I_{h}\right]=E\left[R^{f}\,|\,I_{f}\right]\equiv\mu^{f}\\
     var\left(R^{f}\,|\,I_{h}\right)=var\left(R^{f}\,|\,I_{f}\right)+\eta_{h}\equiv\left(\sigma^{f}\right)^{2}+\eta_{h}
    \end{gather*}
where $\left(\sigma^{h}\right)^{2}$ and $\left(\sigma^{f}\right)^{2}$
are the true variances of domestic and foreign assets, respectively.
Both home and foreign investors share the expected rate of return
of the home (foreign) equity, $\mu^{h}$ and $\mu^{f}$, despite the
different information sets. Nevertheless, as foreign investors make
more precise predictions regarding the rate of return on foreign equities,
the home investors face an additional noise about the foreign equities,
with the amount of $\eta_{h}$, beyond the actual variance of foreign
equity. Likewise, home investors who are familiar with domestic equities
can anticipate the expected return on domestic equities, so foreign
investors face an additional noise with the amount of $\eta_{f}$
beyond the real variance of home equity.

For convenience, we assume that the covariance of domestic-foreign
equity return is constant and independent of information sets, i.e.
$cov\left(R^{h},\,R^{f}\,|\,I_{h}\right)=cov\left(R^{h},\,R^{f}\,|\,I_{f}\right)=\sigma^{hf}$.

From Equation (\ref{ehb-eq2}), assuming an interior solution, the optimal share
of domestic and foreign equities for home investors, $w_{h}^{h}$
and $w_{h}^{f}$, are
    \footnote{Analogously, the foreign investors' optimal shares of home and foreign equities are:
    {\footnotesize{}
        \begin{gather*}
           w_{f}^{h}=\dfrac{1}{\lambda_{f}}\dfrac{\left(\mu^{h}-c_{f}^{h}\right)-\left(\mu^{f}-c_{f}^{f}\right)}{\left(\sigma^{h}\right)^{2}+\left(\sigma^{f}\right)^{2}-2\sigma^{hf}+\eta_{f}}+\dfrac{\left(\sigma^{f}\right)^{2}-\sigma^{hf}}{\left(\sigma^{h}\right)^{2}+\left(\sigma^{f}\right)^{2}-2\sigma^{hf}+\eta_{f}}\\
           w_{f}^{f}=\dfrac{1}{\lambda_{f}}\dfrac{\left(\mu^{f}-c_{f}^{f}\right)-\left(\mu^{h}-c_{f}^{h}\right)}{\left(\sigma^{h}\right)^{2}+\left(\sigma^{f}\right)^{2}-2\sigma^{hf}+\eta_{f}}+\dfrac{\left(\sigma^{h}\right)^{2}-\sigma^{hf}+\eta_{f}}{\left(\sigma^{h}\right)^{2}+\left(\sigma^{f}\right)^{2}-2\sigma^{hf}+\eta_{f}}
        \end{gather*}}}

    \begin{gather} \label{ehb-eq3}
    w_{h}^{h}=\dfrac{1}{\lambda_{h}}\dfrac{\left(\mu^{h}-c_{h}^{h}\right)-\left(\mu^{f}-c_{h}^{f}\right)}{\left(\sigma^{h}\right)^{2}+\left(\sigma^{f}\right)^{2}-2\sigma^{hf}+\eta_{h}}+\dfrac{\left(\sigma^{f}\right)^{2}-\sigma^{hf}+\eta_{h}}{\left(\sigma^{h}\right)^{2}+\left(\sigma^{f}\right)^{2}-2\sigma^{hf}+\eta_{h}}\\
    w_{h}^{f}=\dfrac{1}{\lambda_{h}}\dfrac{\left(\mu^{f}-c_{h}^{f}\right)-\left(\mu^{h}-c_{h}^{h}\right)}{\left(\sigma^{h}\right)^{2}+\left(\sigma^{f}\right)^{2}-2\sigma^{hf}+\eta_{h}}+\dfrac{\left(\sigma^{h}\right)^{2}-\sigma^{hf}}{\left(\sigma^{h}\right)^{2}+\left(\sigma^{f}\right)^{2}-2\sigma^{hf}+\eta_{h}}\nonumber.
    \end{gather}
    
The market-clearing condition is required for equilibrium in the international
financial markets. Let us denote the proportion of global total equity
portfolio owned by a home (foreign) country as $\Pi_{h}\;\left(\Pi_{f}\right)$.
That is, $\Pi_{h}$ and $\Pi_{f}$, can be interpreted as the share
of world wealth belonging to home and foreign countries. The market
clearing condition is,

    \begin{multline} \label{ehb-eq4}
        \Pi_{h}w_{h}^{h}+\underbrace{\left(1-\Pi_{h}\right)}_{\equiv\Pi_{f}}w_{f}^{h}=w^{h*}=\\\Pi_{h}\dfrac{1}{\lambda_{h}}\dfrac{\left(\mu^{h}-c_{h}^{h}\right)-\left(\mu^{f}-c_{h}^{f}\right)}{\left(\sigma^{h}\right)^{2}+\left(\sigma^{f}\right)^{2}-2\sigma^{hf}+\eta_{h}}+\Pi_{h}\dfrac{\left(\sigma^{f}\right)^{2}-\sigma^{hf}+\eta_{h}}{\left(\sigma^{h}\right)^{2}+\left(\sigma^{f}\right)^{2}-2\sigma^{hf}+\eta_{h}}\\
        +\left(1-\Pi_{h}\right)\dfrac{1}{\lambda_{f}}\dfrac{\left(\mu^{h}-c_{f}^{h}\right)-\left(\mu^{f}-c_{f}^{f}\right)}{\left(\sigma^{h}\right)^{2}+\left(\sigma^{f}\right)^{2}-2\sigma^{hf}+\eta_{f}}+\left(1-\Pi_{h}\right)\dfrac{\left(\sigma^{f}\right)^{2}-\sigma^{hf}}{\left(\sigma^{h}\right)^{2}+\left(\sigma^{f}\right)^{2}-2\sigma^{hf}+\eta_{f}}
    \end{multline}
    
    \begin{multline*}
        \Pi_{h}w_{h}^{f}+\underbrace{\left(1-\Pi_{h}\right)}_{\equiv\Pi_{f}}w_{f}^{f}=w^{f*}=\\\Pi_{h}\dfrac{1}{\lambda_{h}}\dfrac{\left(\mu^{f}-c_{h}^{f}\right)-\left(\mu^{h}-c_{h}^{h}\right)}{\left(\sigma^{h}\right)^{2}+\left(\sigma^{f}\right)^{2}-2\sigma^{hf}+\eta_{h}}+\Pi_{h}\dfrac{\left(\sigma^{h}\right)^{2}-\sigma^{hf}}{\left(\sigma^{h}\right)^{2}+\left(\sigma^{f}\right)^{2}-2\sigma^{hf}+\eta_{h}}\\
        +\left(1-\Pi_{h}\right)\dfrac{1}{\lambda_{f}}\dfrac{\left(\mu^{f}-c_{f}^{f}\right)-\left(\mu^{h}-c_{f}^{h}\right)}{\left(\sigma^{h}\right)^{2}+\left(\sigma^{f}\right)^{2}-2\sigma^{hf}+\eta_{f}}+\left(1-\Pi_{h}\right)\dfrac{\left(\sigma^{h}\right)^{2}-\sigma^{hf}+\eta_{f}}{\left(\sigma^{h}\right)^{2}+\left(\sigma^{f}\right)^{2}-2\sigma^{hf}+\eta_{f}}     
    \end{multline*}
where $w^{h*}\;\left(w^{f*}\right)$ is the home's (foreign's) domestic
market capitalization share in the international financial market.
By combining Equations (\ref{ehb-eq3}) and (\ref{ehb-eq4}), 
one can derive the equity home bias from the perspective of a home country, $(w_{h}^{h}-w^{h*})$:

    \begin{multline} \label{ehb-eq5}
    w_{h}^{h}-w^{h*} =\\\left(1-\Pi_{h}\right)\dfrac{1}{\lambda_{h}}\dfrac{\left(\mu^{h}-c_{h}^{h}\right)-\left(\mu^{f}-c_{h}^{f}\right)}{\left(\sigma^{h}\right)^{2}+\left(\sigma^{f}\right)^{2}-2\sigma^{hf}+\eta_{h}}+\left(1-\Pi_{h}\right)\dfrac{\left(\sigma^{f}\right)^{2}-\sigma^{hf}+\eta_{h}}{\left(\sigma^{h}\right)^{2}+\left(\sigma^{f}\right)^{2}-2\sigma^{hf}+\eta_{h}}\\
    -\left(1-\Pi_{h}\right)\dfrac{1}{\lambda_{f}}\dfrac{\left(\mu^{h}-c_{f}^{h}\right)-\left(\mu^{f}-c_{f}^{f}\right)}{\left(\sigma^{h}\right)^{2}+\left(\sigma^{f}\right)^{2}-2\sigma^{hf}+\eta_{f}}-\left(1-\Pi_{h}\right)\dfrac{\left(\sigma^{f}\right)^{2}-\sigma^{hf}}{\left(\sigma^{h}\right)^{2}+\left(\sigma^{f}\right)^{2}-2\sigma^{hf}+\eta_{f}}\;.
    \end{multline}

Based on the property of the variance of two returns, we can easily
verify that $Var(R^{h}-R^{f})=\left(\sigma^{h}\right)^{2}+\left(\sigma^{f}\right)^{2}-2\sigma^{hf}$
is positive. The first two terms of $\left(w_{h}^{h}-w^{h*}\right)$
in Equation (\ref{ehb-eq5}) are related to domestic investors. The first term
explains the relationship between the relative net return on home
equity and the equity home bias. It is sensible that home investors
tend to increase their share of home equities if the net return on
domestic equities is relatively high compared to the one on foreign
equities. The second term shows the relationship between the volatility
of foreign equity and the equity home bias, given the information
frictions of the home country. The latter two terms are associated
with foreign investors' behavior. Foreign investors increase the share
of home equities if home equities give a relatively high net return
compared to foreign equities (the third term). Also, foreign investors
tend to avoid the volatility of foreign equities by purchasing home
equities (the fourth term)
    \footnote{Analogously, a home country increases the share of foreign equity
    if the net return of foreign equities and the risk of home equities
    are relatively higher.}.
    
Moreover, assuming symmetric uncertainty, i.e. $\eta_{h}=\eta_{f}$,
and transaction cost, i.e. $c_{h}^{h}=c_{h}^{f}$ and $c_{f}^{h}=c_{f}^{f}$,
we expect that the equity home bias depends on the correlation between
investors' consumption and return on domestic equities only when the
degree of risk aversion $(\lambda)$ is heterogeneous across countries.
Based on the current framework, consumption is defined as future wealth
from investment. Therefore, we can identify that the correlation between
consumption and domestic return is non-negative, as the optimal shares
of both home and foreign equities are between zero and one. For example,
if there is a higher net return on the domestic equity compared to
the foreign equities, $(\mu^{h}>\mu^{f})$, the equity home bias is
increasing in return on domestic equity only when foreign investors
have a higher degree of risk aversion than domestic investors $(\lambda_{f}>\lambda_{h})$.
This implies that domestic investors face a higher correlation between
consumption and domestic equity return compared to foreign investors.
In such a case, domestic investors tend to increase the domestic equity
share when the domestic equities give a relatively higher return.
Alternatively, if domestic investors are relatively less sensitive
to the portfolio return than foreign investors $(\lambda_{f}<\lambda_{h})$,
the equity home bias is decreasing in return on domestic equities
because foreign investors would increase their domestic equity share
more than domestic investors when a domestic return is higher. When
both domestic and foreign investors have symmetric sensitivity of
their consumption to the return on the portfolio $(\lambda_{f}=\lambda_{h})$,
the equity home bias is independent of the correlation between consumption
and domestic equity return, because the same correlation level between
consumption and domestic return gives both investors to set the same
share of domestic equity. 

To simplify the comparative statics, we can additionally assume
that investors in both countries have the same levels of degree of risk aversion,
i.e., $\lambda_{h}=\lambda_{f}$. Then, the equity home bias described
in Equation (\ref{ehb-eq5}) is simplified without the assumption of  
an identical net return across countries $(\mu^{h}=\mu^{f})$, 
such as Equation (\ref{ehb-eq6}):
    \begin{equation} \label{ehb-eq6}
    w_{h}^{h}-w^{h*} =\left(1-\Pi_{h}\right)\left(w_{h}^{h}-w_{h}^{f}\right)=\left(1-\Pi_{h}\right)\dfrac{\eta}{\left(\sigma^{h}\right)^{2}+\left(\sigma^{f}\right)^{2}-2\sigma^{hf}+\eta}.
    \end{equation}
By taking partial derivatives with respect to either $\eta$
and $\sigma^{hf}$, one can get $\frac{\partial\left(w_{h}^{h}-w^{h*}\right)}{\partial\eta}>0$
and $\frac{\partial\left(w_{h}^{h}-w^{h*}\right)}{\partial\sigma_{hf}}>0$
    \footnote{Alternatively, we can maintain heterogeneity in noise from the information
    frictions across countries as in Equation (\ref{ehb-eq6}). Taking the partial derivative
    with respect to the covariance between home and foreign equities $(\sigma^{hf})$,
    we can get the following form: 
    {\footnotesize{}
    \begin{gather*}
    \frac{\partial\left(w_{h}^{h}-w^{h*}\right)}{\partial\sigma^{hf}}  =\left(1-\Pi_{h}\right)\left[\frac{\eta_{h}}{\left\{ \left(\sigma^{h}\right)^{2}+\left(\sigma^{f}\right)^{2}-2\sigma^{hf}+\eta_{h}\right\} ^{2}}+\frac{\eta_{f}}{\left\{ \left(\sigma^{h}\right)^{2}+\left(\sigma^{f}\right)^{2}-2\sigma^{hf}+\eta_{f}\right\} ^{2}}\right]\\
    +\left(1-\Pi_{h}\right)\left\{ \left(\sigma^{f}\right)^{2}-\left(\sigma^{h}\right)^{2}\right\} \left[\frac{1}{\left\{ \left(\sigma^{h}\right)^{2}+\left(\sigma^{f}\right)^{2}-2\sigma^{hf}+\eta_{h}\right\} ^{2}}-\frac{1}{\left\{ \left(\sigma^{h}\right)^{2}+\left(\sigma^{f}\right)^{2}-2\sigma^{hf}+\eta_{f}\right\} ^{2}}\right]\;.
    \end{gather*}
    }
    Assuming the same variance $(\left(\sigma^{h}\right)^{2}=\left(\sigma^{f}\right)^{2})$,
    the equity home bias is increasing in the covariance of return between
    home and foreign equities.}.  
Given that $\eta$ is positive, a higher correlation between home
and foreign assets leads to a higher equity home bias, which is consistent
with the results of \citet{levy2014home}, in which they argue that higher
substitution between home and foreign equities leads to a higher home
bias because investors do not have incentives to convert their
domestic equities to foreign ones. In the case that the home equity
is a substitute for the foreign equity, i.e., a positive $\sigma^{hf}$,
the existence of information frictions on foreign equity amplifies
the equity home bias.

\section{Empirical Analysis}\label{ehb-empirical}

\subsection{Measures}
As there is no specific data about the equity home bias, this study
measures the equity home bias following the procedure in \citet{coeurdacier2013home}. 
We can define the equity home bias in the home country
as the difference between (A) the proportion of domestic equity in
the total equity portfolio of a home country and (B) the share of
a home country's capitalization in total world equity market capitalization,
i.e., (A) - (B). The following equation (\ref{ehb-eq7}) defines the measure of
equity home bias in this study:

    \begin{equation} \label{ehb-eq7}
    EHB_{h}=1-\dfrac{\frac{HFE_{h}}{TEP_{h}}}{\frac{DMC_{h}}{World\;Total\;Market\;Cap}}
    \end{equation}
where $EHB_{h}$ is the equity home bias, $HFE_{h}$ denotes holding
of foreign equity by the home country, $TEP_{h}$ is the total equity
portfolio in the home country, and $DMC_{h}$ represents domestic
market capitalization. Also, in order to get the total equity
portfolio in the home country $(TEP_{h})$ in
Equation (\ref{ehb-eq7}), the following relationship should be satisfied:

    \begin{equation} \label{ehb-eq8}
    TEP_{h}=DMC_{h}+HFE_{h}-HFE_{f}
    \end{equation}
where $HFE_{f}$ is the foreign holdings of domestic equity. If the
portfolio of the home country is perfectly diversified, $EHB_{h}$
should be zero based on the international capital asset pricing model
(ICAPM) (\citet{coeurdacier2013home}). 

For a measure of unfamiliarity under information frictions about foreign equities, 
this work utilizes the proportion of respondents who answered the survey question,
``How much you trust:
People you meet for the first time?'' from the World Value Survey
(WVS) (\citet{wvs}). Although Hofstede's cultural dimensions data set captures several
aspects of country-specific differences, this is not precisely consistent
with country-specific unfamiliarity about foreign equities. For example,
individualism and uncertainty avoidance in Hofstede's data are not
sufficient alternatives to unfamiliarity 
    \footnote{In Hofstede's model (\url{https://geert-hofstede.com/countries.html}), each
    country's cultural factors are analyzed by 6 dimensions, (1) Power
    Distance, (2) Individualism, (3) Masculinity, (4) Uncertainty Avoidance,
    (5) Long-term Orientation, and (6) Indulgence. \citet{beugelsdijk2010cultural}
    use (2) and (4) for their analysis.}.
    
\subsection{Data}
\subsubsection{Equity Home Bias}

Equity home bias can be calculated using domestic market capitalization
and total portfolio investment across the border with Equations (\ref{ehb-eq7})
and (\ref{ehb-eq8}). The World Bank (WB) and the World Federation of Exchanges
(WFE) provide domestic market capitalization data in terms of market
capitalization in equity. This paper exploits the WB data set because
it provides country-level information
    \footnote{The WFE data are exchange market based, therefore those data do not
    always correspond to the country-based data.}. 
Also, total portfolio investment across the border by nations can
be extracted from the Coordinated Portfolio Investment Survey (CPIS)
of the IMF. In CPIS, there are total portfolio investment assets and
liabilities in equity for each country, which are equivalent to domestic
holdings of foreign equities and foreign holdings of domestic equities.

The period for the empirical analysis in this study is 17 years, from
2001 to 2017. Data on market capitalization earlier than
2001 are available. In contrast, the data for the cross-border investment in equity
in CPIS has been published only since 2001. Data for 2018 are not employed because
of missing values for some countries. To construct a balanced panel
data set, the sample comprises 11 countries (Australia, Canada,
France, Germany, Hong Kong, Japan, South Korea, the Netherlands, Singapore,
Switzerland, and the U.S.) that have reported domestic market capitalization
to WB and cross-border portfolios to CPIS without missing values among
developed countries with the WVS data.

\begin{figure}[ht]
        
        \caption{Equity Home Bias in Several Developed Countries} \label{fig:ehb-homebias}
            \begin{threeparttable}
            
                \begin{centering}
                \includegraphics[scale=0.34]{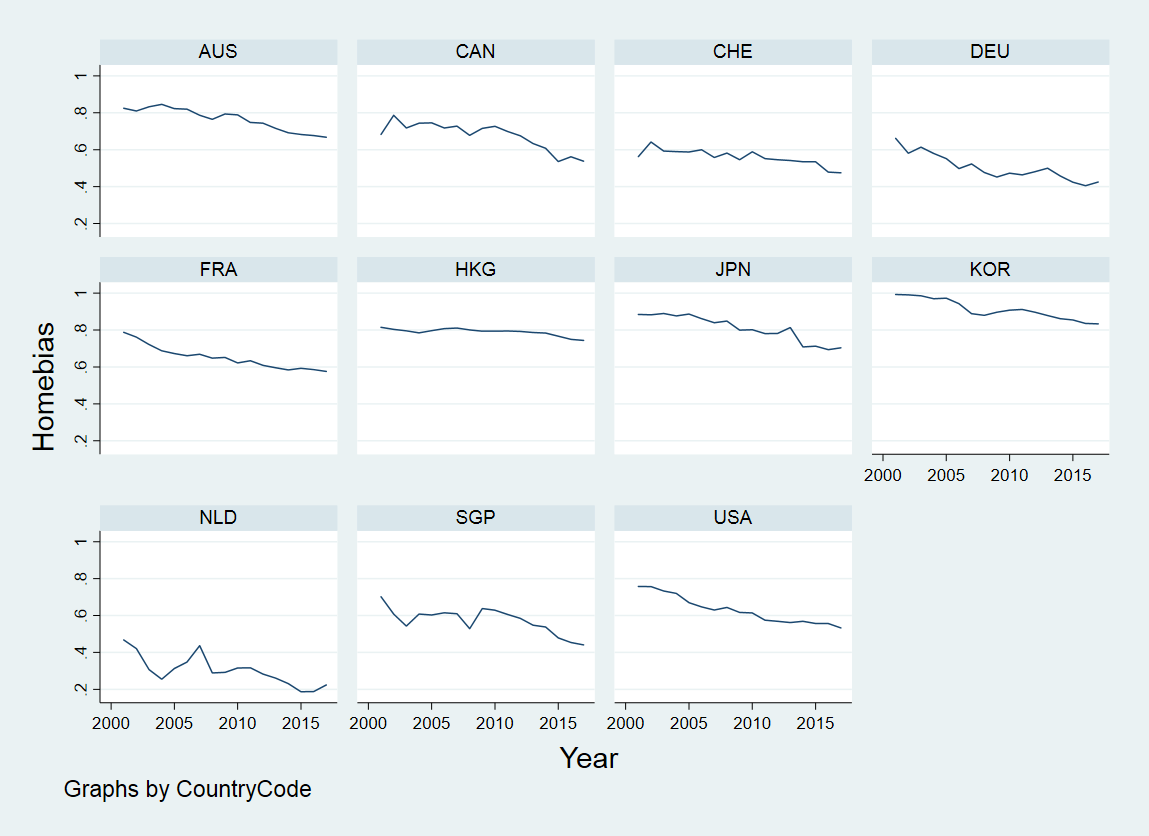}
                \par\end{centering}
                
            \begin{tablenotes}
            \item [Note:] \footnotesize{} Computed from IMF CPIS and World Bank data
            \end{tablenotes}
            \end{threeparttable}
            
            \vspace{0.5 in}
        \end{figure}

All of the selected countries have equity home bias, as shown in Figure
\ref{fig:ehb-homebias}. Most of the non-EU countries, for example, Australia, Canada, Hong
Kong, Japan, South Korea, and the U.S., show a higher equity home
bias compared to the selected EU-countries. For instance, in 2014,
the equity home biases in three non-EU countries were 71\% (Japan),
57\% (U.S.), 53\% (Switzerland) and 60\% (Canada). Although France
has shown temporarily higher home biases than the U.S. and Canada
earlier in the 21st century, other selected countries in the EU, for
example, Germany and the Netherlands, exhibit significantly lower
home equity biases. Capital movement freedom may explain this feature
within EU countries
    \footnote{We could not include Italy into our sample for analysis, 
    because there has been no update regarding
    market capitalization data since 2008.}. 

As shown in Figure \ref{fig:ehb-homebias}, the bias of each country was smaller in 2017
compared to 2001, even if there were fluctuations. The simple average
of equity home bias in these 11 countries was 74\% in 2001, which contrasts
with an average home bias of 56\% in 2017. Considering those eleven
countries are highly developed countries whose financial markets are
deeply integrated, the equity home bias puzzle cannot be explained
only by the transaction cost, including differences in legal systems
and languages.

\subsubsection{Non-fundamental unfamiliarity}
Non-fundamental unfamiliarity is measured from the WVS dataset, survey
data regarding cultural features by nations. WVS has conducted six
waves since 1990; it has interviewed between 1,000 and 2,000
people in each nation per wave. Given this dataset, this paper
considers the proportion of people who respond ``Do not trust much''
or ``Do not trust at all'' to the question: \textit{`` I 'd like
to ask you how much you trust people from various groups. Could you
tell me for each whether you trust people from this group completely,
somewhat, not very much or not at all? - People you meet for the first
time''} in each country
    \footnote{This question is provided only in waves 5 (2005 $\sim$ 2009)
    and 6 (2010 $\sim$ 2014). All of the selected countries,
    except for Hong Kong, Japan and Singapore, reported in wave 5, while only
    eight countries (Australia, Germany, Hong Kong, Japan, South Korea, Netherlands,
    Singapore and the U.S.) reported in wave 6. There are
    no significant changes in the proportion of our interest in the five countries 
    that reported in both waves (Australia, Germany, South Korea, Netherlands, the U.S.). Therefore, we utilize
    data in wave 5 for the measure of the unfamiliarity in each country
    except for Hong Kong, Japan, and Singapore, for which we use wave 6 instead.}.

\begin{figure}[ht]
        
        \caption{The Portion of Respondents for ``Do not trust much'' or ``Do not
        trust at all'' to the Question: \textit{`` I \textquoteleft d like
        to ask you how much you trust people from various groups. Could you
        tell me for each whether you trust people from this group completely,
        somewhat, not very much or not at all? - People you meet for the first
        time.''}} \label{fig:ehb-unfamiliarity}
            \begin{threeparttable}
            
                \begin{centering}
                \includegraphics[scale=0.5]{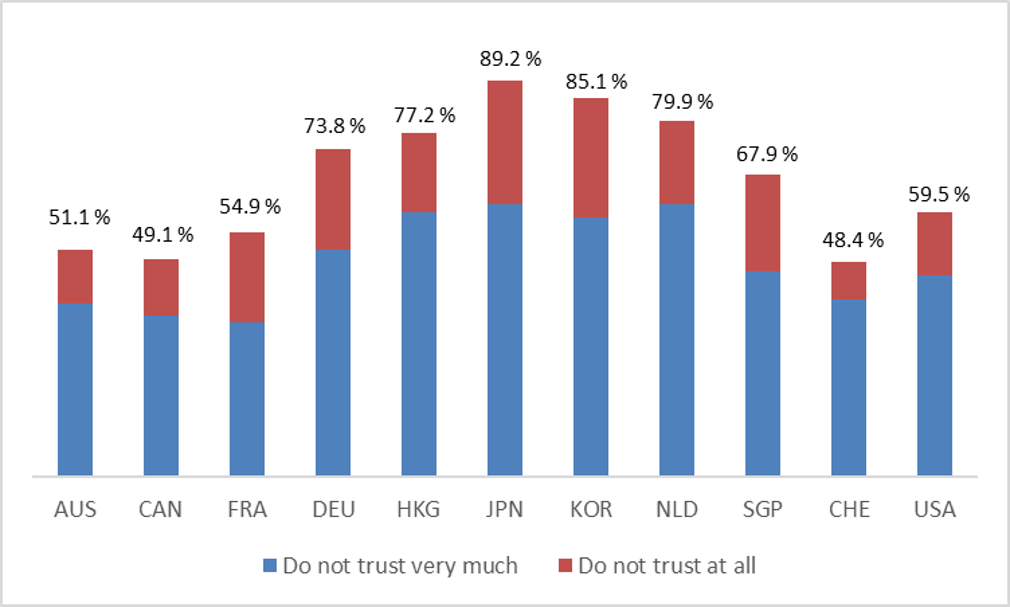}
                \par\end{centering}
                
            \begin{tablenotes}
            \item [Note:] \footnotesize{} Computed from the WVS data
            \end{tablenotes}
            \end{threeparttable}
            
            \vspace{0.5 in}
        \end{figure}

Figure \ref{fig:ehb-unfamiliarity} shows the individual mistrust from unfamiliarity of others
by selected nations. Except for Canada and Switzerland, over half
of respondents in selected countries stated that they do not trust
strangers. Specifically, over 70\% of German, Italian and Japanese
respondents showed mistrust toward unfamiliar people. 

\subsection{Empirical Model and Results}
In the empirical analysis, a panel analysis is implemented. 
As country-specific unfamiliarity is time-invariant in this set up, 
we conduct a feasible generalized linear squared regression (FGLS) method.
Moreover, the FGLS method is useful when the error structure across
the panels is heteroskedastic and correlated.

Our variables of interest are 1) country-level unfamiliarity, 2) the correlation
of return between domestic and foreign equities, and 3) the difference
in annual returns between domestic and foreign equities. Regarding
country-level unfamiliarity, we use the portion of respondents for
``Do not trust much'' or ``Do not trust at all'' to the WVS question:
``Do you trust people you meet for the first time?'' 

For a measure of the correlation of return between domestic and foreign equities, the rate of return on the Global Dow Jones index is assumed as the return on foreign equity. The benchmark equity index in each nation is treated as the equity of the home country
    \footnote{Data were extracted from Yahoo finance and the Wall Street Journal
    online. The country indexes we used are All Ordinaries for Australia, S\&P TSX composite
    index for Canada, CAC 40 for France, DAX for Germany, Hang Seng for
    Hong Kong, Nikkei 225 for Japan, KOSPI for S. Korea, Amsterdam AEX
    index for the Netherlands, Strait Times Index for Singapore, SMI for
    Switzerland, and Dow Jones Industrial Average for the United States.
    Also, the Global Dow Jones index is used to construct the growth rate of
    global equity return. A limitation of this paper is that it does not
    compute the exact covariance of equity return across markets.}. 
    
Based on the daily rate of return on those indexes, we compute the
annual correlation of returns between domestic and foreign equities
for sample countries. The difference in annual returns between domestic
and foreign equities represents the amount of excess return on the
local benchmark index compared to the Global Dow Jones index in each
country
    \footnote{The Global Dow Jones index is a stock index consisting of 150 firms from
    around the world.}.

In terms of control variables, we introduce 1) a Euro currency area
dummy, and 2) a city-level market dummy, and 3) the GDP growth rate. The
Euro currency area dummy stands for free capital movement among the
member countries of the currency union in the European Union (EU). Also, we
introduce a dummy variable that represents city-level markets,
Hong Kong and Singapore. Those financial markets
are highly depend on international financial transactions. Table \ref{tab:ehb-stat} offers
summary statistics for the variables used. A country-level summary
of statistics is attached in Section \ref{ehb-appenC}
    \footnote{It is important to check multicollinearity among independent variables
    in the sample panel dataset. Using the variance inflation factor (VIF),
    we determine that there is no multicollinearity since VIF values from
    independent variables are no more than 2.18, which are much smaller the often-used threshold of 10.}. 

Since there is a decreasing trend in equity home bias in each country,
we control for the trend in equity home bias by adding year as a control variable.
Based on a fixed-effect analysis (FE) by regressing the equity home bias on the
year variable, the estimated coefficient for
the year variable is -0.11 with a standard error of 0.0005, which is significant at the 1\% level.

    \begin{table}
    \begin{threeparttable}
        \caption{Summary Statistics (Aggregate Level)}
        \label{tab:ehb-stat}
    
        \begin{centering}
            \begin{tabular*}{\textwidth}{@{\extracolsep{\fill}}>{\centering}p{1.5in}
            ccccc}
            \toprule 
            {\footnotesize{}Variable} & {\footnotesize{}Mean} & {\footnotesize{}Std. Dev.} & {\footnotesize{}Minimum} & {\footnotesize{}Maximum} & {\footnotesize{}\# of Obs.}\tabularnewline
            \midrule 
            {\footnotesize{}Home Bias} & {\footnotesize{}0.652} & {\footnotesize{}0.173} & {\footnotesize{}0.187} & {\footnotesize{}0.993} & {\footnotesize{}187}\tabularnewline
            {\footnotesize{}Unfamiliarity} & {\footnotesize{}0.668} & {\footnotesize{}0.144} & {\footnotesize{}0.480} & {\footnotesize{}0.890} & {\footnotesize{}11}\tabularnewline
            {\footnotesize{}Correlation with foreign return} & {\footnotesize{}0.582} & {\footnotesize{}0.193} & {\footnotesize{}0.162} & {\footnotesize{}0.869} & {\footnotesize{}187}\tabularnewline
            {\footnotesize{}Difference in return between home and foreign equities} & {\footnotesize{}-0.012} & {\footnotesize{}0.113} & {\footnotesize{}-0.312} & {\footnotesize{}0.355} & {\footnotesize{}187}\tabularnewline
            {\footnotesize{}Hong Kong and Singapore} & {\footnotesize{}-} & {\footnotesize{}-} & {\footnotesize{}0} & {\footnotesize{}1} & {\footnotesize{}11}\tabularnewline
            {\footnotesize{}GDP growth rate} & {\footnotesize{}0.024} & {\footnotesize{}0.024} & {\footnotesize{}-0.056} & {\footnotesize{}0.145} & {\footnotesize{}187}\tabularnewline
            {\footnotesize{}Euro currency area} & {\footnotesize{}-} & {\footnotesize{}-} & {\footnotesize{}0} & {\footnotesize{}1} & {\footnotesize{}11}\tabularnewline
            {\footnotesize{}Year} & {\footnotesize{}-} & {\footnotesize{}-} & {\footnotesize{}2001} & {\footnotesize{}2017} & {\footnotesize{}17}\tabularnewline
            \bottomrule
            \end{tabular*}
    \par
    \end{centering}
    \end{threeparttable}
    \vspace{0.5 in}
    \end{table}

We construct five estimation models to investigate
the relationship of unfamiliarity with the equity
home bias. Model 1 is a benchmark model. The equity home bias is treated as
the dependent variable. Also, unfamiliarity, GDP growth rate, and
the Euro currency area dummy are incorporated as explanatory variables.
In model 2, the correlation of returns between home and global equities
is included to verify whether a higher substitution between domestic
and global equity leads to a higher equity home bias. In model 3,
the correlation between local and global equity returns is replaced
with the difference in return between domestic and global equities
to test whether an excess return on home equities compared to the
global equity has a positive relationship with equity home bias. In
model 4, both the correlation and the difference with the global returns
are considered for the estimation. Finally, in Model 5, we control for
the city-level global financial markets, Hong Kong and Singapore.

    \begin{table}
    \begin{threeparttable}
        \caption{Summary of Empirical Results (Original Equity Home Bias across Sample
        Countries)} \label{tab:ehb-result1}
    
        \begin{centering}
            \begin{tabular*}{\textwidth}{@{\extracolsep{\fill}}cccccc}
            \toprule 
            {\footnotesize{}Variables} & {\footnotesize{}Model 1} & {\footnotesize{}Model 2} & {\footnotesize{}Model 3} & {\footnotesize{}Model 4} & {\footnotesize{}Model 5}\tabularnewline
            \midrule
            \midrule 
            \multirow{2}{*}{{\footnotesize{}Unfamiliarity}} & {\footnotesize{}0.247{*}{*}{*}} & {\footnotesize{}0.074{*}{*}{*}} & {\footnotesize{}0.239{*}{*}{*}} & {\footnotesize{}0.071{*}{*}{*}} & {\footnotesize{}0.110{*}{*}{*}}\tabularnewline
             & {\footnotesize{}(0.010)} & {\footnotesize{}(0.010)} & {\footnotesize{}(0.010)} & {\footnotesize{}(0.013)} & {\footnotesize{}(0.013)}\tabularnewline
            \multirow{2}{*}{{\footnotesize{}Correlation with foreign return}} &  & {\footnotesize{}-0.305{*}{*}{*}} &  & {\footnotesize{}-0.302{*}{*}{*}} & {\footnotesize{}-0.313{*}{*}{*}}\tabularnewline
             &  & {\footnotesize{}(0.018)} &  & {\footnotesize{}(0.021)} & {\footnotesize{}(0.023)}\tabularnewline
            {\footnotesize{}Difference in return b/t} &  &  & {\footnotesize{}0.090{*}{*}{*}} & {\footnotesize{}0.112{*}{*}{*}} & {\footnotesize{}0.100{*}{*}{*}}\tabularnewline
            {\footnotesize{}home and foreign equities} &  &  & {\footnotesize{}(0.010)} & {\footnotesize{}(0.013)} & {\footnotesize{}(0.014)}\tabularnewline
            \multirow{2}{*}{{\footnotesize{}Hong Kong and Singapore}} &  &  &  &  & {\footnotesize{}-0.100{*}{*}{*}}\tabularnewline
             &  &  &  &  & {\footnotesize{}(0.008)}\tabularnewline
            \multirow{2}{*}{{\footnotesize{}GDP growth rate}} & {\footnotesize{}-0.036} & {\footnotesize{}-0.305{*}{*}{*}} & {\footnotesize{}0.046} & {\footnotesize{}-0.211{*}{*}} & {\footnotesize{}0.163}\tabularnewline
             & {\footnotesize{}(0.039)} & {\footnotesize{}(0.090)} & {\footnotesize{}(0.068)} & {\footnotesize{}(0.103)} & {\footnotesize{}(0.102)}\tabularnewline
            \multirow{2}{*}{{\footnotesize{}Euro currency area}} & {\footnotesize{}-0.238{*}{*}{*}} & {\footnotesize{}-0.149{*}{*}{*}} & {\footnotesize{}-0.235{*}{*}{*}} & {\footnotesize{}-0.137{*}{*}{*}} & {\footnotesize{}-0.154{*}{*}{*}}\tabularnewline
             & {\footnotesize{}(0.004)} & {\footnotesize{}(0.006)} & {\footnotesize{}(0.006)} & {\footnotesize{}(0.005)} & {\footnotesize{}(0.006)}\tabularnewline
            \multirow{2}{*}{{\footnotesize{}Year}} & {\footnotesize{}-0.011{*}{*}{*}} & {\footnotesize{}-0.012{*}{*}{*}} & {\footnotesize{}-0.011{*}{*}{*}} & {\footnotesize{}-0.012{*}{*}{*}} & {\footnotesize{}-0.012{*}{*}{*}}\tabularnewline
             & {\footnotesize{}(0.000)} & {\footnotesize{}(0.001)} & {\footnotesize{}(0.000)} & {\footnotesize{}(0.001)} & {\footnotesize{}(0.001)}\tabularnewline
            \multirow{2}{*}{{\footnotesize{}Constant}} & {\footnotesize{}22.034{*}{*}{*}} & {\footnotesize{}24.242{*}{*}{*}} & {\footnotesize{}22.599{*}{*}{*}} & {\footnotesize{}24.649{*}{*}{*}} & {\footnotesize{}24.413{*}{*}{*}}\tabularnewline
             & {\footnotesize{}(0.300)} & {\footnotesize{}(1.029)} & {\footnotesize{}(0.712)} & {\footnotesize{}(1.168)} & {\footnotesize{}(1.045)}\tabularnewline
            \midrule 
            {\footnotesize{}Wald $\chi^{2}$ } & {\footnotesize{}37668.14{*}{*}{*}} & {\footnotesize{}3616.48{*}{*}{*}} & {\footnotesize{}9289.05{*}{*}{*}} & {\footnotesize{}1795.88{*}{*}{*}} & {\footnotesize{}2410.58{*}{*}{*}}\tabularnewline
            {\footnotesize{}(Degree of freedom)} & {\footnotesize{}(4)} & {\footnotesize{}(5)} & {\footnotesize{}(5)} & {\footnotesize{}(6)} & {\footnotesize{}(7)}\tabularnewline
            \bottomrule
            \end{tabular*}
    
    \par
    \end{centering}
    \begin{tablenotes}

        \item [a.] \footnotesize{} *** $p<0.01$, ** $p<0.05$, and * $p<0.1$
        
        \item [b.] \footnotesize{} Estimated coefficients (top) and standard errors (bottom, within parenthesis)

    \end{tablenotes}
    \end{threeparttable}
    \vspace{0.5 in}
    \end{table}

Table \ref{tab:ehb-result1} reports the results from the empirical analysis. 
Among the five estimation models, Model 4 is the preferred specification. Model 4 shows the highest degree of goodness-of-fit in terms of the smallest Wald chi-square statistic, as shown in the last row of Table \ref{tab:ehb-result1}. It implies that Model 4 contains significant explanatory variables necessary in the empirical analysis with the least redundancy associated with an insignificant variable. It also suggests that Model 4 is neither under- nor over-specified compared to the other models.  

We
find a significantly positive relationship of country-level
unfamiliarity with the equity home bias. The coefficient estimates
from Models 1 and 3 in which the correlation of equity returns between
two equities is not included are 0.24. In the case of Models 2 and
4 that incorporate the correlation of returns, the coefficient estimates are 0.07. In model 5 that controls for Hong Kong and Singapore,
the coefficient estimate is 0.1. Even though the point estimates are different
across the empirical models, they are all positive at
a 1\% significance level. In sum, 
the equity home bias is positively related to country-specific unfamiliarity.

As per the correlation between home and foreign equity returns, the results from the empirical analysis show a negative relationship with the equity home bias. The estimates of coefficients are between -0.31 and -0.30 at a 1\% significance level across empirical models. Those results contrast with our hypothesis from the comparative statics in Section \ref{ehb-framework} and the results from \citet{levy2014home}
    \footnote{This counter-intuitive empirical result may be related to several factors, including a difference in the degree of risk aversion $(\lambda_{h}\neq\lambda_{f})$. Moreover, the degree of uncertainty on the foreign equity return from the home country $(\eta_{h})$ is not always equal to the degree of uncertainty on the home equity return from the foreign country $(\eta_{f})$, which also may lead to the counter-intuitive results in this empirical analysis.}.

The difference in return between home and foreign
equities shows a positive relationship with the equity home bias.
The estimated coefficient is between 0.09 and 0.12 at a 1\% significance
level. Those results are opposite to our expectation under the assumption of 
symmetric degree of risk aversion $(\lambda_{f}=\lambda_{h})$ and transaction costs of obtaining
equities, as shown in Equations (\ref{ehb-eq5}) and (\ref{ehb-eq6}) of Section \ref{ehb-framework}. 
The empirical results suggest that home investors acquire home equities more than foreign investors when there is an excess return on home equities compared to foreign ones. Those results are consistent with the ones from the theoretical framework under the assumption that foreign investors have a higher degree risk aversion than domestic investors $(\lambda_{f}>\lambda_{h})$. 

Additionally, we can identify how the equity home
bias is associated with other control variables.
Firstly, we do not find a consistent result regarding the
correlation between the GDP growth rate and the equity home bias.
The coefficient is -0.3 \textasciitilde -0.2 in Models 2 and 4, while
we do not find any significant results from Models 1 and 3. 

A city dummy that represents the city-level independent
financial markets, Hong Kong and Singapore, has a negative
relationship with the equity home bias; the estimate of coefficient
is -0.1 at a 1\% significance level in Model 5. This result is consistent
with the fact that both regions have acted as international financial
hubs in Asia. Since most of the financial transactions associated with
Asian countries depend on international finance in those city-level
markets, the equity home bias in those regions is at a lower level
compared to Korea and Japan.

Furthermore, freedom of capital movement among Euro
currency members (Eurozone) has a negative correlation with the equity
home bias. The empirical results show a negative relationship between 
the Euro currency area dummy and the equity home bias with coefficients
ranging between -0.24 and -0.14 depending on the models. In sum,
this Euro currency area dummy variable can explain a relatively lower
level of home equity bias in the three sample countries whose currency
is the Euro (France, Germany, and the Netherlands), and we can interpret
those results as a lower equity home bias in the Eurozone because
of the free capital movement between member countries.

Regarding a trend of the equity home bias, we determine that the equity
home bias has decreased over time. The estimate of coefficient is
-0.012 $\sim$ -0.011 at a 1\% significance level.

\subsection{Empirical Results with A Transformed Measure of Unfamiliarity}
As reported in Tables \ref{tab:ehb-stat} and \ref{tab:ehb-stat_cty} of the previous subsection, 
the predicted values of the equity home bias are less than 1, which is consistent with 
the support of the primary measure of equity home bias defined in Equation (\ref{ehb-eq7})
, such as $(-\infty,\,1]$. 

In this part, we transformed the measure of the equity home bias with a larger support, $(-\infty,\,+\infty)$, and checked whether the empirical results are consistent with ones from the original data set. Specifically, we defined the latter part in the definition of the equity home bias described in Equation (\ref{ehb-eq7}) as $Q_{h}$, such as: 
    \begin{equation}\label{ehb-eq_a1}
        Q_{h}\equiv\dfrac{\frac{HFE_{h}}{TEP_{h}}}{\frac{DMC_{h}}{World\;Total\;Market\;Cap}}
    \end{equation}
where $Q_{h}$ is the home country's holding portion of foreign equity normalized by the share of its domestic market capitalization share in the world market. As all the values in Equation (\ref{ehb-eq_a1}) are non-negative, $Q_{h}$ should be also non-negative, $(Q_{h}\geq0)$. Then, we took a log-transformation and defined the measure as $g_{h}$ by converting this transformed value into negative, as in Equation (\ref{ehb-eq_a2}):
    \begin{equation}\label{ehb-eq_a2}
        g_{h}\equiv-ln(Q_{h})
    \end{equation}
where the support of $g_{h}$ is ranging from $-\infty$ to $+\infty$. And, the measure $g_{h}$ is positively associated with the equity home bias.

We maintain the previous five estimation models. Therefore, 
a benchmark model, Model 1, focuses only on country-level
unfamiliarity controlling for GDP growth rate and Euro currency area dummy.
We include the correlation of returns between home and global equities
in Model 2, while we replace the correlation of equity returns with
the difference in return between domestic and global equities in Model
3. In model 4, both correlation and difference of returns between
home and global equities are incorporated. Lastly, city-level independent
markets are considered in Model 5.

The results from the modified support of the equity home bias are reported in
Table \ref{tab:ehb-result2a}. It can be observed that the results are consistent with ones in
the previous part that utilizes the original measure of the equity home bias defined in Equation (\ref{ehb-eq7}). 
Among the five estimation models, Model 5 shows the best fit for the empirical analysis in the log-transformed measure case: The Wald chi-square statistic shows the lowest value, 974.77.

    \begin{table}
    \begin{threeparttable}
        \caption{Summary of Empirical Results (Transformed Equity Home Bias
        to the Support between $-\infty$ and $+\infty$, $(-\infty,\,+\infty)$)}
        \label{tab:ehb-result2a}
    
        \begin{centering}
            \begin{tabular*}{\textwidth}{@{\extracolsep{\fill}}cccccc}
            \toprule 
            {\footnotesize{}Variables} & {\footnotesize{}Model 1} & {\footnotesize{}Model 2} & {\footnotesize{}Model 3} & {\footnotesize{}Model 4} & {\footnotesize{}Model 5}\tabularnewline
            \midrule
            \midrule 
            \multirow{2}{*}{{\footnotesize{}Unfamiliarity}} & {\footnotesize{}1.848{*}{*}{*}} & {\footnotesize{}1.292{*}{*}{*}} & {\footnotesize{}1.742{*}{*}{*}} & {\footnotesize{}1.126{*}{*}{*}} & {\footnotesize{}1.292{*}{*}{*}}\tabularnewline
             & {\footnotesize{}(0.035)} & {\footnotesize{}(0.063)} & {\footnotesize{}(0.054)} & {\footnotesize{}(0.076)} & {\footnotesize{}(0.082)}\tabularnewline
            \multirow{2}{*}{{\footnotesize{}Corr. with foreign return}} &  & {\footnotesize{}-1.044{*}{*}{*}} &  & {\footnotesize{}-1.113{*}{*}{*}} & {\footnotesize{}-1.125{*}{*}{*}}\tabularnewline
             &  & {\footnotesize{}(0.064)} &  & {\footnotesize{}(0.084)} & {\footnotesize{}(0.100)}\tabularnewline
            {\footnotesize{}Difference in return b/t} &  &  & {\footnotesize{}0.704{*}{*}{*}} & {\footnotesize{}0.767{*}{*}{*}} & {\footnotesize{}0.624{*}{*}{*}}\tabularnewline
            {\footnotesize{}home and foreign equities} &  &  & {\footnotesize{}(0.071)} & {\footnotesize{}(0.079)} & {\footnotesize{}(0.089)}\tabularnewline
            \multirow{2}{*}{{\footnotesize{}Hong Kong and Singapore}} &  &  &  &  & {\footnotesize{}-0.491{*}{*}{*}}\tabularnewline
             &  &  &  &  & {\footnotesize{}(0.062)}\tabularnewline
            \multirow{2}{*}{{\footnotesize{}GDP growth rate}} & {\footnotesize{}0.917{*}{*}{*}} & {\footnotesize{}0.119} & {\footnotesize{}1.297{*}{*}{*}} & {\footnotesize{}0.189} & {\footnotesize{}2.012{*}{*}{*}}\tabularnewline
             & {\footnotesize{}(0.244)} & {\footnotesize{}(0.260)} & {\footnotesize{}(0.367)} & {\footnotesize{}(0.382)} & {\footnotesize{}(0.508)}\tabularnewline
            \multirow{2}{*}{{\footnotesize{}Euro currency area}} & {\footnotesize{}-0.773{*}{*}{*}} & {\footnotesize{}-0.464{*}{*}{*}} & {\footnotesize{}-0.732{*}{*}{*}} & {\footnotesize{}-0.401{*}{*}{*}} & {\footnotesize{}-0.479{*}{*}{*}}\tabularnewline
             & {\footnotesize{}(0.010)} & {\footnotesize{}(0.019)} & {\footnotesize{}(0.018)} & {\footnotesize{}(0.021)} & {\footnotesize{}(0.024)}\tabularnewline
            \multirow{2}{*}{{\footnotesize{}Year}} & {\footnotesize{}-0.046{*}{*}{*}} & {\footnotesize{}-0.048{*}{*}{*}} & {\footnotesize{}-0.045{*}{*}{*}} & {\footnotesize{}-0.048{*}{*}{*}} & {\footnotesize{}-0.048{*}{*}{*}}\tabularnewline
             & {\footnotesize{}(0.001)} & {\footnotesize{}(0.001)} & {\footnotesize{}(0.002)} & {\footnotesize{}(0.002)} & {\footnotesize{}(0.003)}\tabularnewline
            \multirow{2}{*}{{\footnotesize{}Constant}} & {\footnotesize{}92.941{*}{*}{*}} & {\footnotesize{}97.643{*}{*}{*}} & {\footnotesize{}90.537{*}{*}{*}} & {\footnotesize{}97.140{*}{*}{*}} & {\footnotesize{}97.389{*}{*}{*}}\tabularnewline
             & {\footnotesize{}(2.080)} & {\footnotesize{}(2.743)} & {\footnotesize{}(4.663)} & {\footnotesize{}(4.340)} & {\footnotesize{}(5.641)}\tabularnewline
            \midrule 
            {\footnotesize{}Wald $\chi^{2}$} & {\footnotesize{}18221.04{*}{*}{*}} & {\footnotesize{}3497.20{*}{*}{*}} & {\footnotesize{}2856.06{*}{*}{*}} & {\footnotesize{}1256.26{*}{*}{*}} & {\footnotesize{}974.77{*}{*}{*}}\tabularnewline
            {\footnotesize{}(Degree of freedom)} & {\footnotesize{}(4)} & {\footnotesize{}(5)} & {\footnotesize{}(5)} & {\footnotesize{}(6)} & {\footnotesize{}(7)}\tabularnewline
            \bottomrule
            \end{tabular*}
    \par
    \end{centering}
     \begin{tablenotes}
         \item [a.] \footnotesize{} *** $p<0.01$, ** $p<0.05$, and * $p<0.1$
         \item [b.] \footnotesize{} Estimated coefficients (top) and standard errors (bottom, within parenthesis)
         \item [c.] \footnotesize{} Measure of country-level equity home bias is transformed with the support of $(-\infty,\,+\infty)$ 
    \end{tablenotes}
    \end{threeparttable}
    \vspace{0.5 in}
    \end{table}

As per the country-level unfamiliarity, we find that it has a positive correlation with the equity home bias. The coefficients of estimates range from 1.29 to 1.87 at a 1\% significance level. The estimated coefficients are higher than 1.7 when the empirical model does not incorporate the correlation of returns between two equities. In contrast, the coefficients are smaller when the correlation of equity returns is included in the model. 

In terms of other variables of our interest, we also find that the relationships between the equity home bias and other variables of our interest are empirically consistent with the results obtained by the original equity home bias measure. Regarding the correlation between home and foreign equity returns, the negative estimates of coefficients at a 1\% significance level offers consistency with ones from the empirical analysis in the previous part; they are still opposed to the hypothesis from the comparative statics. The estimated coefficients are between -1.13 and -1.04 at a 1\% significance level across the empirical models. Moreover, the empirical results also show that an excess return on domestic equity positively correlates with the equity home bias. The estimates of coefficients are significantly positive at a 1\% level, ranging from 0.62 to 0.76.

Consistent with the empirical results in the previous part, the equity home bias has a negative relationship with the currency union, such as the Euro area, and the modified measure of the equity home bias is also decreasing over time.

\section{Conclusion}\label{ehb-conclusion}

Equity home bias is relatively stable over time despite a higher
international financial market integration. This study incorporates
information frictions about the return on foreign equity as a factor
causing the equity home bias. We hypothesize that investors depend
on country-specific non-fundamental behavioral factors when there
are information frictions on foreign equities, which raises the equity
home bias. Furthermore, based on the substitutability of equities
across countries, we hypothesize that the correlation of return between
home and foreign equities positively relates to the equity
home bias. 

In the first part of this research, we derive comparative statics
describing relationships between the equity home bias and either 
information frictions or the home-foreign return correlation. Assuming
an identical degree of 1) risk aversion and of 2) uncertainty about
foreign equity return across countries, an analytical relationship
is derived stating that the equity home bias is associated with information frictions
about return on foreign equities. Comparative statics also
hypothesize that the correlation of return between home and foreign equities
raises the equity home bias. In the second part, this work assumes
that portfolios may depend on the unfamiliarity about foreign equities.
In the empirical analysis, we hypothesize
that unfamiliarity positively associates with the equity home
bias. Also, to account for the effect of substitution of equities across
countries, we hypothesize
that correlation of return between home and foreign equities raises
the equity home bias.

Based on a panel consisting of eleven developed countries, the empirical analysis tests the hypotheses described above. We introduce the proportion of respondents to the question, ''Could you tell me for each whether you trust people from this group completely, somewhat, not very much or not at all? - People you meet for the first time'', as as a proxy for the country-specific unfamiliarity from the WVS. For the measure of return correlation return between home and foreign equities, we use an annual correlation of return between the domestic benchmark index and the Global Dow Jones index. We incorporate 1) the difference in return between the domestic benchmark index and Global Dow Jones index, 2) city-level markets such as Hong Kong and Singapore, 3) GDP growth rate, and 4) Euro currency area to the empirical model. We also include time as a regressor because the equity home bias has a decreasing trend.

Setting up five estimation models, we implement an FGLS method allowing heteroskedasticity and correlation across panels. Model 1 considers only country-level unfamiliarity.  Model 2 adds the correlation of return between domestic and foreign equities to Model 1. In contrast, Model 3 replaces this return correlation in Model 2 with the excess return on domestic equity. Model 4 employs all the variables of our interest in Model 4. Model 5 adds a dummy representing the city-level markets in Asia to Model 4.    

The results from the empirical analysis show a positive relationship between the country-level unfamiliarity and the equity home bias. The coefficient estimates range from 0.07 to 0.24. In contrast to \citet{levy2014home}, the coefficient estimates for the correlation of return between home and foreign equities are negative at a 1\% significance level. Besides, an excess return on home equities compared to foreign ones shows a positive correlation with the equity home bias. This positive correlation is consistent with the comparative statics when foreign investors have a higher risk aversion than domestic investors.

To check the robustness of the empirical analysis, we apply a log-transformed measure of the equity home bias to the five estimation models already defined. The empirical results are consistent with the ones using the original measure. Regarding the country-level unfamiliarity, the coefficient estimates are significantly positive, ranging from 1.1 to 1.8. Counter to the  hypothesized theoretical relationships, the correlation of return between domestic and foreign equities shows a negative relationship with the equity home bias; the estimates of coefficients are ranging from -1.1 to -1.0. Moreover, the excess return on domestic equity shows a positive correlation with the equity home bias, which is also opposite to the comparative statics results under the assumption of symmetric degree of risk aversion and transaction costs of purchasing equities.

Both theoretical and empirical results show that country-level unfamiliarity positively relates to the equity home bias. Specifically, in the empirical analysis, the WVS data publicly available is exploited for a measure of unfamiliarity. This work leaves an in-depth exploration with a larger dataset for future research. 
This work accounts for only countries with a highly integrated equity market. Therefore, most developing markets are not included in the sample of empirical analysis. A dataset with extended time-series is expected to provide additional information about the relationship between the country-specific unfamiliarity and the equity home bias.


\newpage
\bibliographystyle{jgea}
\bibliography{ehbbib}

\newpage
\appendices
    \section{Getting a Mean-variance Utility from an Expected Utility Function (Ljungqvist and Sargent (2018))}\label{ehb-appenA}
        
Without loss of generality, a home investor plans to invest his initial
wealth $W_{h}$ in domestic and foreign assets. Given $R_{h}^{P}$
as the random rate of aggregate return on the home investor's portfolio,
we can denote the home investor's future wealth as $\tilde{W}_{h}=W_{h}\left(1+R_{h}^{P}\right)$.
Given $W_{h}$, we can simplify the investor's utility in terms of
his future wealth, $U\left(\tilde{W}_{h}\right)$, as $U\left(R_{h}^{P}\right)$
because $\tilde{W}_{h}$ is fully determined by $R_{h}^{P}$. The
utility function is described in Equation (\ref{ehb-eq9}):
    \begin{equation} \label{ehb-eq9}
    E\left[U\left(R_{h}^{P}\right)\right]=E\left[-e^{-\lambda_{h}R_{h}^{P}}\right].
    \end{equation}
where $\lambda_{h}$ is a parameter of exponential utility function.
From the property of exponential utility function, we can determine
that the parameter in exponential utility function, $\lambda$, is
the Arrow-Pratt index of absolute risk aversion because:
    \begin{equation}  \label{ehb-eq10}
    r.a.=-\frac{U''\left(R_{h}^{P}\right)}{U'\left(R_{h}^{P}\right)}=-\frac{-\lambda_{h}^{2}e^{-\lambda R_{h}^{P}}}{\lambda_{h}e^{-\lambda R_{h}^{P}}}=\lambda_{h}.
    \end{equation}

Next, when we suppose that $R_{h}^{P}$ is drawn from normal distribution
with mean, $\mu_{h}^{P}$, and standard deviation, $\sigma_{h}^{P}$,
the probability density function (PDF), $g(R_{h}^{P})$ is given by:
    \begin{equation}\label{ehb-eq11}
    g(R_{h}^{P})=\frac{1}{\sigma_{h}^{P}\sqrt{2\pi}}e^{-\frac{(R_{h}^{P}-\mu_{h}^{P})^{2}}{2(\sigma_{h}^{P})^{2}}}.
    \end{equation}

Applying this PDF to Equation (\ref{ehb-eq9}), we can expand the expected utility
function such as: 
    \begin{equation} \label{ehb-eq12}
    E\left[U\left(R_{h}^{P}\right)\right]=-\frac{1}{\sigma_{h}^{P}\sqrt{2\pi}}\int_{-\infty}^{\infty}\left[e^{-\left\{ \lambda_{h}R_{h}^{P}+\frac{(R_{h}^{P}-\mu_{h}^{P})^{2}}{2(\sigma_{h}^{P})^{2}}\right\} }\right].
    \end{equation}

We can rearrange the power term in Equation (\ref{ehb-eq13}) such as:
    \begin{equation} \label{ehb-eq13}
    \lambda_{h}R_{h}^{P}+\frac{(R_{h}^{P}-\mu_{h}^{P})^{2}}{2(\sigma_{h}^{P})^{2}}=\frac{\{R_{h}^{P}-\mu_{h}^{P}+\lambda_{h}(\sigma_{h}^{P})^{2}\}^{2}}{2(\sigma_{h}^{P})^{2}}+\lambda_{h}\left(\mu_{h}^{P}-\frac{\lambda_{h}}{2}(\sigma_{h}^{P})^{2}\right).
    \end{equation}
    
Then, we can decompose Equation (\ref{ehb-eq13}) as in Equation (\ref{ehb-eq14}):
    \begin{equation}\label{ehb-eq14}
    E\left[U\left(R_{h}^{P}\right)\right]=-e^{-\lambda_{h}\left(\mu_{h}^{P}-\frac{\lambda_{h}}{2}(\sigma_{h}^{P})^{2}\right)}\frac{1}{\sigma_{h}^{P}\sqrt{2\pi}}\int_{-\infty}^{\infty}\left[e^{-\left\{ \frac{\{R_{h}^{P}-\mu_{h}^{P}+\lambda_{h}(\sigma_{h}^{P})^{2}\}^{2}}{2(\sigma_{h}^{P})^{2}}\right\} }\right].
    \end{equation}
where $\frac{1}{\sigma_{h}^{P}\sqrt{2\pi}}\int_{-\infty}^{\infty}\left[\exp\left(-\left\{ \frac{\{R_{h}^{P}-\mu_{h}^{P}+\lambda_{h}(\sigma_{h}^{P})^{2}\}^{2}}{2(\sigma_{h}^{P})^{2}}\right\} \right)\right]=1$
because it is the area over the entire support when the mean is $\mu_{h}^{P}-\lambda_{h}(\sigma_{h}^{P})^{2}$
and the standard deviation is $\sigma_{h}^{P}$. The simplified form
of the expected utility function is     
    \begin{equation} \label{ehb-eq15}
    E\left[U\left(R_{h}^{P}\right)\right]=-e^{-\lambda_{h}\left(\mu_{h}^{P}-\frac{\lambda}{2}(\sigma_{h}^{P})^{2}\right)}.
    \end{equation}

Hence, maximization of $E[U\left(\tilde{W}_{h}\right)]$ is equivalent
to maximizing the following quadratic expression:

    \begin{equation} \label{ehb-eq16}
    E\left[\hat{U}\left(R_{h}^{P}\right)\right]=\mu_{h}^{P}-\frac{\lambda_{h}}{2}(\sigma_{h}^{P})^{2}
    \end{equation}
where $\lambda_{h}$ measures the degree of home investor's risk
aversion.

        \pagebreak
    \section{A Proxy for Country-specific Unfamiliarity}\label{ehb-appenB}
        
    \begin{table}[!ht]
    \begin{threeparttable}
        \caption{The Portion of Respondents for ``Do not trust much'' or ``Do not
        trust at all'' to the Question: `` I \textquoteleft d like to ask
        you how much you trust people from various groups. Could you tell
        me for each whether you trust people from this group completely, somewhat,
        not very much or not at all? - People you meet for the first time.''}
        \label{tab:ehb-unfamiliarity}
    
        \begin{centering}
            \begin{tabular*}{\textwidth}{@{\extracolsep{\fill}}cccc}
            \toprule 
            \multirow{2}{*}{{\footnotesize{}Country}} & \multicolumn{2}{c}{{\footnotesize{}Unfamiliarity}} & \multirow{2}{*}{{\footnotesize{}Sum}}\tabularnewline
            \cmidrule{2-3} \cmidrule{3-3} 
             & {\footnotesize{}Do not trust much} & {\footnotesize{}Do not trust at all} & \tabularnewline
            \midrule 
            {\footnotesize{}Australia} & {\footnotesize{}38.9\%} & {\footnotesize{}12.2\%} & {\footnotesize{}51.1\%}\tabularnewline
            {\footnotesize{}Canada} & {\footnotesize{}36.3\%} & {\footnotesize{}12.8\%} & {\footnotesize{}49.1\%}\tabularnewline
            {\footnotesize{}France} & {\footnotesize{}34.8\%} & {\footnotesize{}20.1\%} & {\footnotesize{}54.9\%}\tabularnewline
            {\footnotesize{}Germany} & {\footnotesize{}51.1\%} & {\footnotesize{}22.7\%} & {\footnotesize{}73.8\%}\tabularnewline
            {\footnotesize{}Hong Kong} & {\footnotesize{}59.5\%} & {\footnotesize{}17.8\%} & {\footnotesize{}77.2\%}\tabularnewline
            {\footnotesize{}Japan} & {\footnotesize{}61.3\%} & {\footnotesize{}27.9\%} & {\footnotesize{}89.2\%}\tabularnewline
            {\footnotesize{}S. Korea} & {\footnotesize{}58.4\%} & {\footnotesize{}26.7\%} & {\footnotesize{}85.1\%}\tabularnewline
            {\footnotesize{}Netherlands} & {\footnotesize{}61.2\%} & {\footnotesize{}18.7\%} & {\footnotesize{}79.9\%}\tabularnewline
            {\footnotesize{}Singapore} & {\footnotesize{}46.2\%} & {\footnotesize{}21.6\%} & {\footnotesize{}67.9\%}\tabularnewline
            {\footnotesize{}Switzerland} & {\footnotesize{}39.8\%} & {\footnotesize{}8.6\%} & {\footnotesize{}48.4\%}\tabularnewline
            {\footnotesize{}United States} & {\footnotesize{}45.5\%} & {\footnotesize{}14.0\%} & {\footnotesize{}59.5\%}\tabularnewline
            \bottomrule
            \end{tabular*}
    \par
    \end{centering}
    \end{threeparttable}
    \vspace{0.5 in}
    \end{table}

        \pagebreak
    \section{Country-level Summary of Statistics}\label{ehb-appenC}
        
    \begin{table} [!h]
    \begin{threeparttable}
        \caption{Summary Statistics (Country Level)}
        \label{tab:ehb-stat_cty}
    
        \begin{centering}
            \begin{tabular}{>{\raggedright}p{0.8in}>{\centering}p{2in}cccc}
            \toprule 
            {{\footnotesize{}Country}} & {\footnotesize{}Variable} & {\footnotesize{}Mean} & {\footnotesize{}Std. Dev.} & 
            {{\footnotesize{}Minimum}} & {\footnotesize{}Maximum}\tabularnewline
            \midrule
            \midrule 
            \multirow{4}{0.8in}{{\footnotesize{}Australia}} & {\footnotesize{}Home Bias} & {\footnotesize{}0.766} & {\footnotesize{}0.060} & {\footnotesize{}0.668} & {\footnotesize{}0.846}\tabularnewline
             & {\footnotesize{}Correlation with foreign return} & {\footnotesize{}0.348} & {\footnotesize{}0.094} & {\footnotesize{}0.168} & {\footnotesize{}0.507}\tabularnewline
             & {\footnotesize{}Difference in return between home and foreign equities} & {\footnotesize{}-0.016} & {\footnotesize{}0.092} & {\footnotesize{}-0.157} & {\footnotesize{}0.181}\tabularnewline
             & {\footnotesize{}GDP growth rate} & {\footnotesize{}0.029} & {\footnotesize{}0.007} & {\footnotesize{}0.019} & {\footnotesize{}0.041}\tabularnewline
            \midrule
            \multirow{4}{0.8in}{{\footnotesize{}Canada}} & {\footnotesize{}Home Bias} & {\footnotesize{}0.676} & {\footnotesize{}0.075} & {\footnotesize{}0.536} & {\footnotesize{}0.787}\tabularnewline
             & {\footnotesize{}Correlation with foreign return} & {\footnotesize{}0.648} & {\footnotesize{}0.094} & {\footnotesize{}0.476} & {\footnotesize{}0.769}\tabularnewline
             & {\footnotesize{}Difference in return between home and foreign equities} & {\footnotesize{}-0.015} & {\footnotesize{}0.086} & {\footnotesize{}-0.165} & {\footnotesize{}0.159}\tabularnewline
             & {\footnotesize{}GDP growth rate} & {\footnotesize{}0.020} & {\footnotesize{}0.015} & {\footnotesize{}-0.029} & {\footnotesize{}0.032}\tabularnewline
            \midrule
            \multirow{4}{0.8in}{{\footnotesize{}France}} & {\footnotesize{}Home Bias} & {\footnotesize{}0.651} & {\footnotesize{}0.062} & {\footnotesize{}0.576} & {\footnotesize{}0.788}\tabularnewline
             & {\footnotesize{}Correlation with foreign return} & {\footnotesize{}0.775} & {\footnotesize{}0.055} & {\footnotesize{}0.663} & {\footnotesize{}0.850}\tabularnewline
             & {\footnotesize{}Difference in return between home and foreign equities} & {\footnotesize{}-0.045} & {\footnotesize{}0.074} & {\footnotesize{}-0.180} & {\footnotesize{}0.066}\tabularnewline
             & {\footnotesize{}GDP growth rate} & {\footnotesize{}0.012} & {\footnotesize{}0.013} & {\footnotesize{}-0.029} & {\footnotesize{}0.028}\tabularnewline
            \midrule
            \multirow{4}{0.8in}{{\footnotesize{}Germany}} & {\footnotesize{}Home Bias} & {\footnotesize{}0.504} & {\footnotesize{}0.072} & {\footnotesize{}0.405} & {\footnotesize{}0.662}\tabularnewline
             & {\footnotesize{}Correlation with foreign return} & {\footnotesize{}0.769} & {\footnotesize{}0.049} & {\footnotesize{}0.702} & {\footnotesize{}0.867}\tabularnewline
             & {\footnotesize{}Difference in return between home and foreign equities} & {\footnotesize{}0.004} & {\footnotesize{}0.095} & {\footnotesize{}-0.202} & {\footnotesize{}0.182}\tabularnewline
             & {\footnotesize{}GDP growth rate} & {\footnotesize{}0.013} & {\footnotesize{}0.022} & {\footnotesize{}-0.056} & {\footnotesize{}0.041}\tabularnewline
            \midrule
            \multirow{4}{0.8in}{{\footnotesize{}Hong Kong}} & {\footnotesize{}Home Bias} & {\footnotesize{}0.790} & {\footnotesize{}0.020} & {\footnotesize{}0.744} & {\footnotesize{}0.815}\tabularnewline
             & {\footnotesize{}Correlation with foreign return} & {\footnotesize{}0.424} & {\footnotesize{}0.093} & {\footnotesize{}0.264} & {\footnotesize{}0.573}\tabularnewline
             & {\footnotesize{}Difference in return between home and foreign equities} & {\footnotesize{}0.000} & {\footnotesize{}0.156} & {\footnotesize{}-0.261} & {\footnotesize{}0.243}\tabularnewline
             & {\footnotesize{}GDP growth rate} & {\footnotesize{}0.037} & {\footnotesize{}0.029} & {\footnotesize{}-0.025} & {\footnotesize{}0.087}\tabularnewline
            \bottomrule
            \end{tabular}
    \par
    \end{centering}
    \end{threeparttable}
    \vspace{0.5 in}
    \end{table}
    
    \begin{table}
    \begin{threeparttable}
        \begin{centering}
            \begin{tabular}{>{\raggedright}p{0.8in}>{\centering}p{2in}cccc}
            \toprule 
            {{\footnotesize{}Country}} & {\footnotesize{}Variable} & {\footnotesize{}Mean} & {\footnotesize{}Std.  Dev.} & 
            {{\footnotesize{}Minimum}} & {\footnotesize{}Maximum}\tabularnewline
            \midrule
            \midrule 
            \multirow{4}{0.8in}{{\footnotesize{}Japan}} & {\footnotesize{}Home Bias} & {\footnotesize{}0.810} & {\footnotesize{}0.070} & {\footnotesize{}0.694} & {\footnotesize{}0.890}\tabularnewline
             & {\footnotesize{}Correlation with foreign return} & {\footnotesize{}0.388} & {\footnotesize{}0.094} & {\footnotesize{}0.184} & {\footnotesize{}0.520}\tabularnewline
             & {\footnotesize{}Difference in return between home and foreign equities} & {\footnotesize{}-0.008} & {\footnotesize{}0.137} & {\footnotesize{}-0.275} & {\footnotesize{}0.303}\tabularnewline
             & {\footnotesize{}GDP growth rate} & {\footnotesize{}0.008} & {\footnotesize{}0.020} & {\footnotesize{}-0.054} & {\footnotesize{}0.042}\tabularnewline
            \midrule 
            \multirow{4}{0.8in}{{\footnotesize{}S. Korea}} & {\footnotesize{}Home Bias} & {\footnotesize{}0.912} & {\footnotesize{}0.054} & {\footnotesize{}0.834} & {\footnotesize{}0.993}\tabularnewline
             & {\footnotesize{}Correlation with foreign return} & {\footnotesize{}0.375} & {\footnotesize{}0.099} & {\footnotesize{}0.162} & {\footnotesize{}0.521}\tabularnewline
             & {\footnotesize{}Difference in return between home and foreign equities} & {\footnotesize{}0.049} & {\footnotesize{}0.142} & {\footnotesize{}-0.187} & {\footnotesize{}0.355}\tabularnewline
             & {\footnotesize{}GDP growth rate} & {\footnotesize{}0.038} & {\footnotesize{}0.017} & {\footnotesize{}0.007} & {\footnotesize{}0.074}\tabularnewline
            \midrule 
            \multirow{4}{0.8in}{{\footnotesize{}Netherlands}} & {\footnotesize{}Home Bias} & {\footnotesize{}0.302} & {\footnotesize{}0.081} & {\footnotesize{}0.187} & {\footnotesize{}0.468}\tabularnewline
             & {\footnotesize{}Correlation with foreign return} & {\footnotesize{}0.774} & {\footnotesize{}0.054} & {\footnotesize{}0.651} & {\footnotesize{}0.846}\tabularnewline
             & {\footnotesize{}Difference in return between home and foreign equities} & {\footnotesize{}-0.049} & {\footnotesize{}0.075} & {\footnotesize{}-0.177} & {\footnotesize{}0.100}\tabularnewline
             & {\footnotesize{}GDP growth rate} & {\footnotesize{}0.013} & {\footnotesize{}0.018} & {\footnotesize{}-0.037} & {\footnotesize{}0.038}\tabularnewline
            \midrule
            \multirow{4}{0.8in}{{\footnotesize{}Singapore}} & {\footnotesize{}Home Bias} & {\footnotesize{}0.573} & {\footnotesize{}0.069} & {\footnotesize{}0.441} & {\footnotesize{}0.702}\tabularnewline
             & {\footnotesize{}Correlation with foreign return} & {\footnotesize{}0.443} & {\footnotesize{}0.099} & {\footnotesize{}0.255} & {\footnotesize{}0.594}\tabularnewline
             & {\footnotesize{}Difference in return between home and foreign equities} & {\footnotesize{}-0.018} & {\footnotesize{}0.132} & {\footnotesize{}-0.205} & {\footnotesize{}0.282}\tabularnewline
             & {\footnotesize{}GDP growth rate} & {\footnotesize{}0.052} & {\footnotesize{}0.038} & {\footnotesize{}-0.011} & {\footnotesize{}0.145}\tabularnewline
            \midrule
            \multirow{4}{0.8in}{{\footnotesize{}Switzerland}} & {\footnotesize{}Home Bias} & {\footnotesize{}0.560} & {\footnotesize{}0.042} & {\footnotesize{}0.475} & {\footnotesize{}0.642}\tabularnewline
             & {\footnotesize{}Correlation with foreign return} & {\footnotesize{}0.706} & {\footnotesize{}0.075} & {\footnotesize{}0.503} & {\footnotesize{}0.796}\tabularnewline
             & {\footnotesize{}Difference in return between home and foreign equities} & {\footnotesize{}-0.040} & {\footnotesize{}0.095} & {\footnotesize{}-0.198} & {\footnotesize{}0.136}\tabularnewline
             & {\footnotesize{}GDP growth rate} & {\footnotesize{}0.018} & {\footnotesize{}0.015} & {\footnotesize{}-0.022} & {\footnotesize{}0.041}\tabularnewline
            \midrule
            \multirow{4}{0.8in}{{\footnotesize{}United States}} & {\footnotesize{}Home Bias} & {\footnotesize{}0.630} & {\footnotesize{}0.075} & {\footnotesize{}0.533} & {\footnotesize{}0.758}\tabularnewline
             & {\footnotesize{}Correlation with foreign return} & {\footnotesize{}0.756} & {\footnotesize{}0.076} & {\footnotesize{}0.609} & {\footnotesize{}0.869}\tabularnewline
             & {\footnotesize{}Difference in return between home and foreign equities} & {\footnotesize{}-0.001} & {\footnotesize{}0.120} & {\footnotesize{}-0.312} & {\footnotesize{}0.173}\tabularnewline
             & {\footnotesize{}GDP growth rate} & {\footnotesize{}0.019} & {\footnotesize{}0.015} & {\footnotesize{}-0.025} & {\footnotesize{}0.038}\tabularnewline
            \bottomrule
            \end{tabular}
    \par
    \end{centering}
    \end{threeparttable}
    \vspace{0.5 in}
    \end{table}


\end{document}